\input harvmac
% \draftmode
\noblackbox
%-------------------------
% This paper uses harvmac
%-------------------------
\font\ticp=cmcsc10
 
\def\Title#1#2{\rightline{#1}\ifx\answ\bigans\nopagenumbers\pageno0\vskip1in
\else\pageno1\vskip.8in\fi \centerline{\titlefont #2}\vskip .5in}

\font\ticp=cmcsc10
\font\ttsmall=cmtt10 at 8pt

%
%-------------------
%  definitions
%-------------------

%

%\def\S{\Sigma}

\def\({\left (}
\def\){\right )}
%
%-------------------
% references
\lref\bgm{P. Breitenlohner, D. Maison, and G. Gibbons, Commun. Math. Phys.
{\bf 120} (1988) 295.}
\lref\hms{ G. Horowitz, J. Maldacena and A. Strominger, hep-th/9603109.}
\lref\ms{J. Maldacena and A. Strominger, hep-th/9603060; C. Johnson,
R. Khuri, R. Myers, hep-th/9603061.}
\lref\spn{J. Breckenridge, R. Myers, A. Peet and C. Vafa, hep-th/9602065.}
\lref\ascv{A. Strominger and C. Vafa, hep-th/9601029.}
\lref\cama{C. Callan and J. Maldacena, hep-th/9602043.}
\lref\ghas{G. Horowitz and A. Strominger, hep-th/9602051.}
\lref\cvetd{M. Cvetic and D. Youm, hep-th/9508058; hep-th/9512127.}
\lref\koll{R. Kallosh and B. Koll, hep-th/9602014; R. Dijkgraaf,
E. Verlinde and H. Verlinde, hep-th/9603126.}
\lref\vbd{J. Breckenridge, D. Lowe, R. Myers, A.  Peet, A. Strominger and
C. Vafa, hep-th/9603078.}
\lref\cveticnew{ M. Cvetic and D. Youm, hep-th/9603147.}
\lref\sugra{E. Cremmer and B. Julia, Nucl. Phys. {\bf B159}
(1979) 141; B. De Wit and H. Nicolai, Nucl. Phys. {\bf B208}
(1982) 323.}
\lref\harstro{J.A. Harvey and A. Strominger, Nucl.
Phys. {\bf B449} (1995) 535.}
%-------------------
% title page
%-------------------
%
\baselineskip 12pt
\Title{\vbox{\baselineskip12pt
\line{\hfil  PUPT-1608, UCSBTH-96-06}
\line{\hfil \tt hep-th/9603195} }}
{\vbox{
{\centerline{Statistical Entropy of Nonextremal}}
{\centerline{Four-Dimensional Black Holes and U-Duality}}
}}
\centerline{\ticp Gary T. Horowitz$^{\dagger}$,\footnote{}{\ttsmall
gary@cosmic.physics.ucsb.edu, lowe@tpau.physics.ucsb.edu,
malda@puhep1.princeton.edu} David A. Lowe$^{\dagger}$ and Juan M.
Maldacena$^{\natural }$}
\bigskip
\vskip.1in
\centerline{$^\ddagger$\it Department of Physics, University of California,
Santa Barbara, CA 93106, USA}
\vskip.1in
\centerline{$^{\natural}$\it Joseph Henry Laboratories, Princeton
University,
Princeton, NJ 08544, USA}
\bigskip
\centerline{\bf Abstract}
We identify the states in string theory which are responsible for the
entropy of near-extremal rotating four-dimensional black holes in
$N=8$ supergravity. For black holes
far from extremality (with no rotation), the Bekenstein-Hawking
entropy is exactly
matched
by a mysterious duality invariant
extension of the formulas derived for near-extremal black holes states.

\Date{March, 1996}

Recent developments in string theory have led, for the first time,
to an understanding of black hole entropy from a microscopic point
of view. In \ascv\  it was shown that the Bekenstein-Hawking entropy of an
extremal, nonrotating,
five-dimensional black hole precisely counts the number of
BPS states in string theory with the given charges (in the limit of large
charges). This agreement has since then
 been extended in a  number of directions.
Extreme 5D black holes with rotation \spn , extreme 4D
black holes \ms , and slightly nonextreme 5D black holes \refs{\cama ,
 \ghas ,\vbd }
 have all been shown
to have a Bekenstein-Hawking entropy which agrees with the number of
corresponding states
in string theory. One goal of the present work is to show that this
agreement continues to hold for slightly nonextremal 4D black holes.

The restriction to extreme or near-extreme black holes arises since we
can only count states at weak coupling, while black holes only exist at strong
coupling. For extremal  configurations, one can argue that
interactions are absent on the basis of supersymmetry
 and the extrapolation of the number
of states from
weak to strong coupling is justified.
For near-extremal black holes there are situations
in which the interactions
are again suppressed.
For black holes far from extremality,
there appears to be no reason why a weakly coupled description is
applicable.

Nevertheless, it was shown in \hms\  that there is a sense in
which even black holes far from extremality can be thought of as composed
of weakly interacting fundamental objects in string theory.
The objects one needs are
the same  ones which yield the states of extremal black holes:
extended solitons known as D-branes and fundamental strings. More precisely,
ref. \hms\  considered a class of five-dimensional black holes labeled by the
energy, three charges, and
the asymptotic values of two scalars. The three charges are carried by
onebranes, fivebranes and strings (or anti-branes, which are just branes with
the opposite orientation and with  the opposite sign of the charge).
One can replace the original six parameters in
the solution by the number of branes, anti-branes and strings
($N_1,\ N_{\bar 1},\ N_5,\ N_{\bar 5},\ n_R, \ n_L$) by matching
the energy, three gauge charges and two scalar charges of
these noninteracting objects with that
of the black hole. In terms of these new
variables the black hole entropy takes the suggestive form
\eqn\smira{S= 2 \pi( \sqrt{ N_1} + \sqrt{  N_{ \bar 1}})
( \sqrt{ N_5} + \sqrt{  N_{ \bar 5}})( \sqrt{ n_L} + \sqrt{ n_R})~.}
This expression applies to all black holes, even those which are far
from extremality.  The symmetry of this expression is consistent with
U-duality which permutes the three types of fundamental objects.
It was argued \hms\ that \smira\ arises naturally from counting states in
string theory, in the sense that it correctly reproduces the number of
string states in three different weak-coupling limits, and is the simplest
duality invariant expression with this property.
However,  no derivation of the general formula directly from counting
states in string
theory is currently available.

Since the significance of the above expression for the black hole entropy
is not yet well understood,
it is important to know whether it is special to five dimensions
 or if it  applies more generally. In this paper we
will show that the entropy of four-dimensional black holes can be expressed
in a form directly analogous to \smira . In \ms\  it was shown that the
states of extremal
four-dimensional black holes can be described in terms of
D-twobranes,
solitonic fivebranes, D-sixbranes, and open strings.
We will consider the nonextremal version of these  solutions
which is  an eight parameter
family of four-dimensional black holes.
By comparing the mass, gauge charges, and scalar charges (which are
pressures in the internal directions)
of the black hole with those of a set of  {\it noninteracting}
branes and anti-branes, we will rewrite the Bekenstein-Hawking entropy
in a form analogous to \smira .
We will show that in certain limits (corresponding to near-extremal black
holes),
the entropy formula we obtain indeed represents the number of states of this
collection of branes at weak coupling.

The generalization of these black hole solutions to include rotation has
recently been found \cveticnew.  We will show that
 in the limit of small rotation and near
extremality
the black hole entropy again agrees with the number of
string states.

We will be considering Type II string theory
compactified on $T^6 = T^4 \times S^1 \times \hat S^1$, which
gives $N=8$ supergravity in four dimensions.
In ref. \cvetd\ general spherically symmetric
black hole solutions of $N=4$ supergravity
in four dimensions were considered. Using these solutions it
is straightforward to construct the  general class
of black holes  in $N=8$ supergravity.
The starting point for this construction is a
solution with four nonzero $U(1)$ gauge fields (carrying two electric and
two magnetic charges) and three nontrivial scalars \cvetd.
The Einstein metric is
\eqn\metric{
\eqalign{
ds^2 &= - f^{-1/2}(r)\(1 - {r_0  \over r}\) dt^2 +
 {f^{1/2}(r) }\left[
\left( 1 -{r_0 \over r} \right)^{-1}  dr^2
+  r^2 (d\theta^2 + \sin^2\theta d\phi^2)\right]~,\cr
f(r) &=
\(1 +{ r_0 \sinh^2
\alpha_2  \over r}\)\(1 + { r_0  \sinh^2 \alpha_5 \over r}\)
\(1+ { r_0  \sinh^2 \alpha_6 \over r}\)\(1 + { r_0
\sinh^2 \alpha_p \over r }\)~.\cr}
}
This metric is parameterized by the
five  independent quantities $\alpha_2$, $\alpha_5$,
$\alpha_6$, $\alpha_p$ and $r_0$.
 The event horizon lies at $r=r_0$. The special case $\alpha_2 = \alpha_5
 =\alpha_6=\alpha_p$ corresponds to the Reissner-Nordstr\"om metric.
The overall solution contains three additional parameters which are related
to the asymptotic values of the three scalars. From the ten-dimensional
viewpoint, these are the
volume of the
4-torus $(2\pi)^4 V$, and the  radii of $S^1$ and $\hat S^1$, $R_1$ and $R_2$.

The physical charges are expressed in terms of these
quantities as
\eqn\charges{
\eqalign{
Q_2  &= {  r_0 V \over g }
\sinh 2\alpha_2 ~, \cr
Q_5 &= { r_0  R_2   }
\sinh 2\alpha_5 ~, \cr
Q_6 &= {  r_0 \over  g}
\sinh 2\alpha_6 ~, \cr
n &=  {  r_0 V R_1^2 R_2 \over g^2 }
\sinh 2\alpha_p ~, \cr}
}
where $g$ is the ten-dimensional string coupling and we have chosen conventions
such that $\alpha'=1$ and
the four-dimensional Newton constant is $G_4 =  g^2/(8 V R_1 R_2)$.
Note that in these conventions the
string coupling is such that  $g \rightarrow 1/g$ under S duality.

The ADM mass of the solution is
\eqn\admmass{
M= { r_0  V R_1 R_2 \over g^2}
 (\cosh 2\alpha_2+
\cosh 2\alpha_5 + \cosh 2 \alpha_6 + \cosh 2 \alpha_p )
}
and the Bekenstein-Hawking entropy is
\eqn\bhentropy{
S ={A\over 4G_4} = {8 \pi r_0^2 V R_1 R_2 \over g^2} \cosh \alpha_2
\cosh \alpha_5 \cosh \alpha_6 \cosh \alpha_p ~.
}

There are three nontrivial scalar fields present
in the solution and
associated with these scalar fields
are three pressures (scalar charges)
\eqn\scharge{
\eqalign{
P_1 &= { r_0 V  R_1 R_2 \over g^2} (\cosh 2\alpha_2 +
\cosh 2 \alpha_6 - \cosh 2\alpha_5 - \cosh 2\alpha_p) ~,\cr
P_2 &= { r_0 V  R_1 R_2 \over g^2} (\cosh 2 \alpha_2-
\cosh 2\alpha_6) ~,\cr
P_3 &= { r_0 V  R_1 R_2 \over g^2} (\cosh 2\alpha_5 -
\cosh 2\alpha_p ) ~.\cr}
}

In the ten-dimensional theory, the four charges \charges\ are carried
by twobranes, fivebranes, sixbranes and strings.
The D-sixbranes  wrap  around $T^4\times S^1\times
\hat S^1$, the solitonic fivebranes wrap around
$T^4 \times S^1$ and the D-twobranes  wrap
around $S^1\times \hat S^1$. The strings carry momentum
along the $S^1$ direction.
In the spirit of \hms\ we calculate the
values for the mass and scalar charges
of each type of brane or string.
This can be calculated from the solution we have presented by
taking the four extremal limits: $r_0 \rightarrow 0, \
\alpha_i \rightarrow \pm \infty$ with $Q_i$ and $\alpha_j \ (j\ne i)$ fixed.
We find that D-twobranes have mass and pressures
\eqn\twobranes{
M=P_1 =P_2 = { R_1 R_2 \over g }~,~~~~~~~~
P_3 = 0 ~,}
while for the sixbranes we have
\eqn\sixbranes{
M = P_1 = -P_2 = { V  R_1 R_2  \over g }~,~~~~~~~~~P_3 =0~.
}
For the solitonic fivebrane we have
\eqn\fivebrane{
M=  -P_1 = P_3 = { V R_1 \over g^2 }~,~~~~~~~~~~~~P_2 =0 ~,}
and for the momentum we find
\eqn\momentum{
M= -P_1 = -P_3 = { 1 \over R_1}~, ~~~~~~~~~~~~~P_2=0~.}

Using these relations plus the charges  \charges\  we
trade in the eight parameters of the
solution for the eight quantities $(n_R, n_L,
N_2,  N_{\bar 2}, N_5,  N_{\bar 5}, N_6,  N_{\bar 6})$
which are the numbers of
right(left)-moving momentum
modes, twobranes, anti-twobranes, fivebranes,
anti-fivebranes, sixbranes and anti-sixbranes.
We do this by matching the mass \admmass ,
pressures \scharge ,
and gauge charges \charges\ with those of a collection of
noninteracting branes. This leads to
\eqn\branenum{
\eqalign{
n_R & = {r_0  V R_1^2 R_2 \over 2 g^2} e^{2\alpha_p}~,  \cr
N_2 &= {r_0  V  \over 2 g} e^{2\alpha_2}~, \cr
N_5 &= {r_0  R_2 \over 2}
e^{2\alpha_5 }~,  \cr
N_6 &= {r_0  \over 2 g} e^{2\alpha_6}~, \cr}
\qquad
\eqalign{
n_L  &= {r_0 V R_1^2 R_2 \over 2 g^2} e^{-2\alpha_p}~, \cr
 N_{\bar 2} &= {r_0  V  \over 2 g} e^{-2\alpha_2}~, \cr
 N_{\bar 5} &= {r_0 R_2 \over 2 }e^{-2\alpha_5}~, \cr
 N_{\bar 6} &= {r_0    \over 2 g} e^{-2\alpha_6}~. \cr}
}

In terms of the brane numbers, the ADM mass is
reexpressed as
\eqn\badm{
M = {1\over R_1} (n_R+n_L)+ {R_1 R_2 \over g} (N_2+
 N_{\bar 2}) + {V R_1\over g^2}(N_5+ N_{\bar 5}) +
{V R_1 R_2 \over g}(N_6+ N_{\bar 6})~,
}
the gauge charges are simply differences of the
brane numbers, and the other parameters are
\eqn\vollen{
V= \sqrt{ {N_2  N_{\bar 2} }\over {N_6  N_{\bar 6} }} ,
\qquad R_2 = \sqrt{ {{N_5  N_{\bar 5}}\over {g^2 N_6
 N_{\bar 6}}}}, \qquad R_1^2 R_2 = \sqrt{ {{g^2 n_R n_L}
\over {N_2  N_{\bar 2} }}}~.
}

The entropy \bhentropy\ then takes the surprisingly simple form
\eqn\entropyb{
S = 2\pi (\sqrt{n_R} +\sqrt{n_L})(\sqrt{N_2}+\sqrt{ N_{\bar 2} })
(\sqrt{N_5}+
\sqrt{ N_{\bar 5}})(\sqrt{N_6}+\sqrt{ N_{\bar 6} })~.
}
This is the analog of \smira\ for four-dimensional black holes. When
one term in each factor vanishes, the black hole is extremal. In this
case, \entropyb\ agrees with the number of bound states of these
branes at weak coupling \ms.  Although we cannot derive the general
formula from counting string states, we can do so in certain limits
corresponding to near-extremal black holes.
Consider the case when $ N_{\bar 2} =  N_{\bar 5} =  N_{\bar 6}
=0$ and  $R_1$ is large. We see from \badm\ that the lightest excitations
will be the momentum modes.
The extremal limit is obtained by also setting the number of left movers
to zero $n_L =0$. In that case the entropy can be calculated
 \ms\  as the entropy of a one-dimensional gas of
$ 4  N_2  N_5 N_6 $ bosonic particles plus an equal number of
fermionic particles with total energy $E = n_R /R_1 $, which
gives $S = 2 \pi \sqrt{ N_2 N_5 N_6 n_R} $.
In the near-extremal limit we also include left movers, which will be
noninteracting if $R_1$ is large.  Hence the entropy will be the sum
\eqn\nearlong{
S = 2 \pi \sqrt{ N_2 N_5 N_6 } ( \sqrt{n_R} +  \sqrt{n_L} )~
}
which clearly agrees with \entropyb\ when $ N_{\bar 2} \sim  N_{\bar 5} \sim
N_{\bar 6} \sim 0 $. Note that these antibrane
excitations are very massive when
$R_1$ is large, so one can see from \branenum , \badm\ that
their number will be very small in the near-extremal limit
 and their contribution to the entropy will be negligible.
We could do a similar calculation for the cases in which the lightest
particles are the other branes. Since U-duality interchanges the
different branes and strings, one expects a result similar to \nearlong\
with the indices permuted. Equation \entropyb\ is clearly the simplest
duality invariant expression which agrees with these different nonextremal
limits.

We have considered only four types of charges.
 Reducing Type II string theory to four
dimensions on $T^6$ leads to a theory whose low energy limit is $N=8$
supergravity. This contains 28 gauge fields and 70 scalars. The gauge fields
can carry either electric or magnetic charges, so there are 56 possible
charges. Each of these charges is carried by a different type of soliton
in the ten-dimensional theory. From black hole uniqueness theorems
\bgm\ it is clear that the Bekenstein-Hawking entropy of the  general solution
depends on the energy
and 56 ``solution generating parameters" that add charge.
However, these parameters are not the physically normalized charges,
but also involve the asymptotic values of the scalars.
 From the
 special form of the
coupling of scalars to gauge fields in $N=8$
supergravity \sugra, one sees that a basis may be chosen
for the scalars in which only 56 of them enter in the normalization
of the
gauge charges.
One can view these parameters as 55 scalars and the total energy. The
entropy can then be viewed as
a function of 56+56 parameters
which may be interpreted as the number of solitons and anti-solitons.

Since the full theory should be $E_7$ invariant we should
be able to write the general entropy formula in an invariant way.
If we denote by $V^A_1$ the 56-dimensional vector giving the number of
solitons and by
$ V^A_2 $ the number of anti-solitons, the formula for the
entropy may take the form
\eqn\eseven{
S = 2 \pi \sum_{i,j,k,l} \sqrt{ T_{ABCD} V^A_i V^B_j V^C_k V^D_l }~,
}
where $T_{ABCD}$ is the quartic invariant considered in  \koll , where
this formula was derived for the extremal case ($ V^A_2 = 0$).

We now consider adding rotation to the black holes discussed above.
Since the rotation dependent terms in the solution fall off faster
at infinity than the charges, the definition of the brane numbers
\branenum\ is unchanged.
If we again take nearly extremal black holes with
$ N_{\bar 2} \sim  N_{\bar 5} \sim   N_{\bar 6} \sim 0$,
 and $R_1$ large,
the Bekenstein-Hawking entropy  takes the form \cveticnew \footnote{
$^1$}{There is a difference in the definition of $J$ from \cveticnew ,
here we are measuring $J$ in units of $\hbar$.}
\eqn\neent{
S=2\pi  \( \sqrt{ n_R N_2 N_5 N_6} +
\sqrt{ n_L N_2 N_5 N_6 - J^2 } \)~.
}
where $J$ is the angular momentum of the black hole.
This agrees precisely with the counting of string states as follows. With $R_1$
much larger than
the other compact dimensions and
with just twobranes and sixbranes present,
the D-brane excitations of this system are described
by a 1+1-dimensional field theory which turns out to be a
$(4,4)$ superconformal sigma model \spn.
The fivebrane breaks the right-moving supersymmetry
\harstro,
leaving us with $(0,4)$ superconformal symmetry.
The $N=4$ superconformal algebra gives rise to
a left-moving  $SU(2)$ symmetry. Since fermionic
states in the sigma model become spinors in spacetime, the action
of $O(3)$ spatial rotations has a natural
action on this $SU(2)$. The charge $F_L$ under
one  $U(1)$ subgroup of this $SU(2)$ will then be
related to the four-dimensional angular
momentum (along one of the three axes)
 carried by the left movers by $J= F_L /2 $.
Due to the presence of the fivebrane the right-moving $SU(2)$
symmetry of the original $(4,4)$ superconformal field theory is broken and
the right movers cannot carry macroscopic angular
momentum.
The number of states with fixed
$n_L$, $n_R$, $F_L\gg 1$ may be computed as in
\refs{\spn,\vbd} to yield the entropy
\eqn\dent{
S= 2 \pi \sqrt{{ c\over  6 }} ( \sqrt{n_R} + \sqrt{\tilde n_L})
{}~,
}
where $\tilde n_L = n_L-
6 J^2/ c$ is the effective number of left movers that
one is free to change once one has demanded that
we have a given macroscopic angular momentum.
For our problem the central
charge is $c=6 N_2 N_5 N_6$ \ms, thus the
entropy \neent\ agrees with the D-brane formula
\dent.

It is interesting to take the extremal limit of these rotating
black holes, when the mass takes the minimum value consistent with
given angular momentum and charges. This happens when
$\tilde n_L =0 $, so the left movers are constrained to just carry the angular
momentum and do not contribute
to the entropy.
When the angular momentum is nonzero, even the extremal black hole is not
supersymmetric.
Using \dent\ and writing the result in
terms of the charge $n = n_R - n_L $ we find
\eqn\ehawken{
S=2\pi \sqrt{J^2 + n Q_2 Q_5 Q_6}~,
}
which indeed agrees with the entropy of an extremal
charged rotating black hole \cveticnew .
Notice the surprising fact that although  we derived this formula
in the large $R_1$ regime (and  $J/M^2 \ll 1$), it continues to
be valid for arbitrary values of the parameters.
Since this is far from the BPS state, we had no reason to expect the
weak-coupling counting to continue to agree  with the black hole
entropy.\footnote{$^\dagger$}{
In five dimensions, taking the extremal limit of the results in
\vbd\ one obtains the microscopic entropy
$S = 2 \pi \sqrt {Q_1 Q_5 n + J_1 J_2 } $. This again agrees with the
entropy of a black hole with two rotation parameters. This black hole is
supersymmetric only when $J_1 = -  J_2$ (and also for $J_1 = J_2 $ with
the opposite sign of one of the charges). }

To summarize, we  first considered four-dimensional nonrotating black holes.
We argued that there is a sense in which one can view
the general nonextremal black hole as composed of a collection of
noninteracting branes and anti-branes. The number of branes of each type
is determined by matching physical properties of the branes with those
of the black hole.
In terms of these numbers, the Bekenstein-Hawking entropy takes the
simple form \entropyb.
We were able to show that in certain limits, this expression agrees with
the number of states of this collection of branes and anti-branes at weak
coupling. A complete derivation of these formula remains an outstanding
challenge. We also showed that for nearly extremal {\it rotating}
black holes,
the entropy again agrees with the number of string states.
Surprisingly, the extremal rotating black hole entropy was precisely
matched by a D-brane counting argument, even far beyond
the regime in which this counting  was done.

There have been earlier indications \cama\ that the counting of string
states at weak coupling agrees with the black hole entropy even in situations
where one could not justify the extrapolation to strong coupling.
We found another example of this
in the case of extremal rotating black holes. The surprising success of these
weak-coupling arguments indicates that understanding black hole entropy
may be even simpler than it appears today.

\vskip 1cm

{\bf Acknowledgements}

\vskip .5cm

We would like to thank M. Cvetic, R. Kallosh, B. Koll, A. Peet, A. Strominger,
and L. Susskind
for useful discussions.
The research of  G.H. is supported in part by NSF Grant PHY95-07065, D.L is
supported in part by NSF
Grant PHY91-16964, and
J. M. is supported in part by DOE grant
DE-FG02-91ER40671.

\listrefs
\end